\documentclass[floats,preprint,prb,aps]{revtex4}
\usepackage{graphicx}
\begin{document}
%\baselineskip 20.75pt
%=========================================================================

\vspace{0.5cm} \preprint{{\em Submitted to Phys.~Rev.~{\bf B}}
\hspace{3.0in} Web galley}
\date{\today}
\vspace{2.7in}

\title{Collective properties of indirect excitons in coupled quantum
wells in random field}

\author{Oleg L. Berman$^{1}$, Yurii E. Lozovik$^{2}$,  David W.
Snoke$^{3}$, and Rob D. Coalson$^{1}$}

\affiliation{\mbox{$^{1}$ Department of Chemistry, University of
Pittsburgh,}   \\ Pittsburgh, PA 15260, USA  \\
\mbox{$^{2}$ Institute of Spectroscopy, Russian Academy of
Sciences,}  \\ 142190 Troitsk, Moscow Region, Russia \\
\mbox{$^{3}$Department of Physics and Astronomy, University of
Pittsburgh,}  \\ 3941 O'Hara Street, Pittsburgh, PA 15260 USA  }

%=========================================================================

\date{\today}

\vspace{2.7in}

\begin{abstract}
The influence of a random field induced by impurities, boundary
irregularities etc. on the superfluidity of a
quasi-two-dimensional  ($2D$)  system of spatially indirect
excitons in coupled quantum wells is studied. The interaction
between excitons is taken into account in the ladder
approximation. The random field is allowed to be large compared to
the dipole-dipole repulsion between excitons. The coherent
potential approximation (CPA) allows us to derive the exciton
Green's function for a wide range of the random field, and the CPA
results are used in the weak-scattering limit, which results in
the second-order Born approximation. The Green's function of the
collective excitations for the cases of (1) equal electron and
hole masses and  (2) the ``heavy hole'' limit are derived
analytically. For quasi-two-dimensional excitonic systems, the
density of the superfluid component and the Kosterlitz-Thouless
temperature of the superfluid phase transition are obtained, and
are found to decrease as the random field increases. This puts
constraints on the experimental efforts to observe excitonic
superfluidity.

\vspace{0.2 in}



\end{abstract}

\maketitle {}

%\newpage
%-----------------------------------------------------------------------
%-------------------------------------------------------------------------
%-------------------------------------------------------------------------

\section{Introduction}

Superfluidity in a system of spatially indirect excitons in
coupled quantum wells (CQW) has been predicted by Lozovik and
Yudson,\cite{Lozovik} and several subsequent theoretical
studies\cite{Klyuchnik,Shevchenko,Lerner,Dzjubenko,Kallin,Knox,
Yoshioka,Birman,Littlewood,Vignale,Berman,Berman_Tsvetus,Berman_Willander,Ulloa}
  have suggested that this should be
manifested as persistent electric currents, quasi-Josephson
phenomena and unusual properties in strong magnetic fields. In the
past ten years, a number of experimental studies have focused on
observing these
behaviors.\cite{Snoke_paper,Snoke_paper_Sc,Chemla,Krivolapchuk,Timofeev,Zrenner,Sivan,Snoke}
The coupled quantum well system is conceptually simple: negative
electrons are trapped in a two-dimensional plane, while an equal
number of positive holes is trapped in a parallel plane a distance
$D$ away. One of the appeals of this system is that the electron
and hole wavefunctions have very little overlap, so that the
excitons can have very long lifetime  ($> 100$ ns), and therefore
they can be treated as metastable particles to which
quasiequlibrium statistics apply. Also when $D$ is large enough,
the interactions between the excitons are entirely dipole-dipole
repulsive.

Much of the theory of excitonic systems has not taken into account
the role of disorder, which is created by impurities and boundary
irregularities of the quantum wells. In real experiments, however,
disorder plays a very important role. Although the inhomogeneous
broadening linewidth of typical GaAs-based samples has been
improved from around 20 meV to less than 1 meV,\cite{Snoke_paper}
the disorder energy is still not much smaller than the
exciton-exciton repulsion energy. At a typical exciton density of
a few times $10^{10}$ cm$^{-2}$, the interaction energy of the
excitons is several meV.  On the other hand, the typical disorder
energy of 1 meV is low compared to the typical exciton binding
energy of 5 meV.  Typical thermal energies at liquid helium
temperatures are $k_BT = 0.2 - 2$ meV.

If the chemical potential of the interacting exciton system
(controlled by the characteristic dipole-dipole interaction at
fixed exciton density) is smaller than the characteristic disorder
energy, the exciton system is expected to be localized in separate
lakes in the minima of the random potential created by the
disorder. The Bose condensation in this case is similar to that
for Bose atoms in trap
\cite{Kim,Anderson,Ensher,Ketterle,Ketterle_Miesner,Daflovo} (for
the role of exciton tunneling, etc., see Ref.
[\onlinecite{Lozovik_tbp}]). In the opposite limit, one has a
translationally invariant, extended exciton system with a random
potential field. In the latter case, the critical properties and
quasi-long non-diagonal order are similar to the case without
disorder (see, e.g., Ref. [\onlinecite{Pomirchy}] and references
cited therein), but the disorder suppresses the
Kosterlitz-Thouless transition temperature and the superfluid
density.

Earlier studies of disorder in exciton systems included theory of
the transport properties of direct and indirect excitons and
magnetoexcitons in random fields,\cite{Ruvinsky_ps} the influence
of various random fields on excitonic and magnetoexcitonic
absorption of light,\cite{Ruvinsky_jetp,Ablyazov} and Anderson
localization of excitons.\cite{Gevorkyan}  The effect of a weak
random field on the collective properties and superfluidity of
excitons in nonuniform systems was analyzed in
Ref.[\onlinecite{Berman_Ruvinsky}], including the cases of a
dilute gas of three-dimensional (3D) excitons, two-dimensional
(2D) excitons in a single quantum well, and indirect excitons in
coupled quantum wells in a random field. In that work the random
field was assumed to be much smaller than the exciton-exciton
interaction. In two-dimensional systems, the excitonic interaction
in the Bogoliubov approximation is valid only in non-physically
low densities because of the divergence of the two-dimensional
scattering amplitude in the Born approximation.\cite{Yudson}
Therefore, the ladder approximation must be used at low densities
to treat properly the interaction between two-dimensional
excitons.\cite{Yudson,Abrikosov} In
Ref.[\onlinecite{Berman_Ruvinsky}], the contribution to the
exciton Green's function from the interaction of the excitons with
the random field was derived by perturbation theory, which limited
the strength of the random field that could be studied. In this
paper we study the case of a random field which is not necessarily
small compared to the dipole-dipole repulsion between excitons.
The coherent potential approximation  (CPA) allows us to derive
the 2D indirect exciton Green's function for a wide range of
random field strengths, resulting in the second order Born
approximation in the weak scattering limit  (the second order Born
approximation Green's function for 3D direct excitons was obtained
by Gevorkyan and Lozovik\cite{Gevorkyan}). We predict that in the
low-temperature limit, the density $n_{s}$ of the superfluid
component in CQW systems and the temperature of the superfluid
transition (the Kosterlitz-Thouless temperature
$T_{c}$\cite{Kosterlitz}) are decreasing functions of the random
field.

A typical example of a two-dimensional system of weakly
interacting bosons is a system of indirect excitons in coupled
quantum wells
(GaAs/AlGaAs)\cite{Snoke_paper,Snoke_paper_Sc,Chemla,Krivolapchuk,Timofeev,Zrenner,Sivan}.
Fluctuations of the thickness of a quantum well, which arise
during the fabrication process,  impurities in the system, and
disorder in the alloy of the barriers can all lead to the
appearance of a random field. Of these, spectral analysis of the
exciton luminescence shows that alloy disorder, with a
characteristic length scale short compared to the excitonic Bohr
radius of around 100~\AA, plays the most important role.

The paper is organized in the following way.  In Sec.~\ref{single}
the Green's function of the single exciton in the random field is
derived analytically in the coherent potential approximation (CPA)
in the weak-scattering limit (resulting in the second order Born
approximation) for the cases of (1) equal electron and hole masses
and (2) the ``heavy hole'' limit.  In
Sec.~\ref{collective_spectrum} the Green's function and the the
spectrum of collective excitations for a system of dilute indirect
excitons in a random field are obtained within the ladder
approximation. In Sec.~\ref{KT_f_tr}  the dependencies of the
density of the superfluid component and the temperature of the
Kosterlitz-Thouless transition on the random field are derived. In
Sec.~\ref{discussion} we we present our conclusions and discuss
possible experimental manifestations of superfluidity of indirect
excitons in CQWs in a random field.

%-------------------------------------------------------------------------
%-------------------------------------------------------------------------

\section{The Green's function of the single exciton in the random field}
\label{single}

In our model, the random potential $V(\mathbf{r})$ acting on an
electron and hole is considered to be delta function correlated
Gaussian noise, such that
%-------------------------------------------------------------------------
\begin{eqnarray}
\label{gaussian} \left\langle V(\mathbf{r})V(\mathbf{r}')
\right\rangle = g \delta^{d}(\mathbf{r} - \mathbf{r}'),
\hspace{0.5in} \left\langle V(\mathbf{r}) \right\rangle = 0 ,
\end{eqnarray}
%-------------------------------------------------------------------------
where $d$ is the dimensionality of the space (for the spatially
indirect exciton in CQWs $d = 2$). An
electron is subjected to the potential $V_{e} = \alpha_{e}
V(\mathbf{r})$ and a hole to $V_{h} = - \alpha_{h} V(\mathbf{r})$,
where $\alpha_{e}$ and $\alpha_{h}$ are constants.

We consider the characteristic length of the random field
potential  $l$ to be much lower than the average distance between
excitons $r_{s} \sim 1/\sqrt{\pi n}$  ($l \ll 1/\sqrt{\pi n}$,
where $n$ is the total exciton density). Therefore, in order to
obtain the Green's function of the excitons with dipole-dipole
repulsion in the random field,  we first obtain the Green's
function of a single exciton in the random field  (not interacting
with other excitons), and then apply perturbation theory with
respect to the dipole-dipole repulsion between excitons, using the
system of the non-interacting excitons as a reference system.

The Green's function of the center of mass of the isolated exciton
at $T = 0$ in the momentum-frequency domain ($G^{(0)}(\mathbf{p},
\omega)$) in the random field in the coherent potential
approximation (CPA) is given by\cite{Gevorkyan} (here and below
$\hbar = 1$)
%-------------------------------------------------------------------------
\begin{eqnarray}
\label{green_0} G^{(0)}(\mathbf{p}, \omega) = \frac{1}{\omega -
\varepsilon_{0}(p) + \mu + i Q(\mathbf{p}, \omega) }  ,
\end{eqnarray}
%-------------------------------------------------------------------------
where $\mu$ is the chemical potential of the system, and
$\varepsilon_{0}(p) = p^{2}/2M$ is the spectrum of the center mass
of the exciton in the ``clean'' system ($M = m_{e} + m_{h}$ is the
mass of the exciton; $m_{e}$ and $m_{h}$ are the electron and hole
masses, respectively). The function $Q(\mathbf{p}, \omega)$ is
determined by the effective random field acting on the center of
mass of the exciton. For zero random field, $Q(\mathbf{p}, \omega)
\rightarrow 0$. If $g M \ll E_{b}$, where $E_{b}$ is the binding
energy of the indirect exciton ($E_{b} \sim e^{2}/\epsilon D$ at
$D \gg \rho (D)$\cite{Nishanov}, $D$ being the distance between
$e$ and $h$ wells; $\rho (D) = (8a)^{1/4}D^{3/4}$ is the radius of
the 2D indirect exciton at $D \gg a$, $\epsilon$ is the dielectric
constant, $a = \epsilon/4m_{e-h}e^{2}$ is the two-dimensional
excitonic Bohr radius in the limit $D \rightarrow 0$; $e$ is the
electron charge, and $m_{e-h} = m_{e}m_{h}/(m_{e} + m_{h})$), then
the function $Q(\mathbf{p}, \omega)$ in the coherent potential
approximation is given by\cite{Gevorkyan}
%-------------------------------------------------------------------------
\begin{eqnarray}
\label{Q_def1} Q (\mathbf{p}, \omega) = \frac{1}{2} \int
G^{(0)}(\mathbf{q}, \omega) B(|\mathbf{p} - \mathbf{q}|)
\frac{d^{d}q}{(2\pi)^{d}}  ,
\end{eqnarray}
%-------------------------------------------------------------------------
where
%----------------------------------------------------------------------
\begin{equation}\label{fourier}
    B(\mathbf{p}) \equiv  \int d^{d} r B(\mathbf R) e^{- i \mathbf{p R}},
\end{equation}
%----------------------------------------------------------------------
with $\mathbf{R} = \mathbf{R}_{1} -  \mathbf{R}_{2}$; in the
coordinate domain $B(\mathbf{R}_{1},\mathbf{R}_{2})$ has the form
\cite{Gevorkyan} (it will be shown below that
$B(\mathbf{R}_{1},\mathbf{R}_{2})$ depends only on $\mathbf{R} =
\mathbf{R}_{1} - \mathbf{R}_{2}$)
%----------------------------------------------------------------------
\begin{eqnarray}\label{B_def}
    &&
    B(\mathbf{R}_{1},\mathbf{R}_{2}) = g \int d^{d}
    r \nonumber \\
    &&  \times \left\{  \alpha_{e}\left( \frac{M}{m_{h}}\right)^{d}
\left| \varphi_{0}
    \left((\mathbf{r} - \mathbf{R}_{1})\frac{M}{m_{h}} \right)\right|^{2} -
    \alpha_{h}\left( \frac{M}{m_{e}}\right)^{d} \left| \varphi_{0}
    \left((\mathbf{R}_{1} - \mathbf{r})\frac{M}{m_{e}}
    \right)\right|^{2}\right\}    \nonumber \\
     &&  \times \left\{  \alpha_{e}\left( \frac{M}{m_{h}}\right)^{d}
\left| \varphi_{0}
    \left((\mathbf{r} - \mathbf{R}_{2})\frac{M}{m_{h}} \right)\right|^{2} -
    \alpha_{h}\left( \frac{M}{m_{e}}\right)^{d} \left| \varphi_{0}
    \left((\mathbf{R}_{2} - \mathbf{r})\frac{M}{m_{e}}
    \right)\right|^{2}\right\}  ,
\end{eqnarray}
%----------------------------------------------------------------------
where $\varphi_{0} (r)$ is the ground-state wave function of an
exciton, corresponding to the relative motion of the electron and
hole ($\mathbf{r} = \mathbf{r}_{e} - \mathbf{r}_{h}$). For a
two-dimensional  indirect exciton with spatially separated
electron and hole at large interwell distances $D$ ($D \gg a$)
   the ground-state wave function is given by\cite{Nishanov}
%-------------------------------------------------------------------------
\begin{eqnarray}
\label{indir_exciton}  \varphi_{0} (r) = \frac{1}{\pi \rho^{2}(D)}
\exp\left( -\frac{r^{2}}{2\rho^{2}(D)}\right) .
\end{eqnarray}
%-------------------------------------------------------------------------

Substituting $\varphi_{0} (r)$ from Eq.~(\ref{indir_exciton}) into
Eq.~(\ref{B_def}), and assuming $d = 2$,  we obtain
%-------------------------------------------------------------------------
\begin{eqnarray}
\label{B_r}  B(R) &=& \frac{2(\alpha_{e} -
\alpha_{h})^{2}g}{\pi^{3}\rho^{2}} \exp\left(
-\frac{2R^{2}}{\rho^{2}}\right) , \hspace{0.5in} m_{e} = m_{h} ;
\nonumber \\
B(R) &=& \frac{\alpha_{e}^{2} g}{2 \pi^{3}\rho^{2}} \exp\left(
-\frac{R^{2}}{\rho^{2}}\right) , \hspace{0.5in} m_{h} \gg m_{e}; \
\  \alpha_{h} \left( \frac{m_{h}}{m_{e}} \right)^{2} \ll
\alpha_{e}.
\end{eqnarray}
%-------------------------------------------------------------------------
Using Eq.~(\ref{fourier}), we obtain the Fourier transform of
$B(R)$
%-------------------------------------------------------------------------
\begin{eqnarray}
\label{B_p}  B(p) &=& \frac{(\alpha_{e} -
\alpha_{h})^{2}g}{4\pi^{4}} \exp\left(
-\frac{\rho^{2}p^{2}}{8}\right) , \hspace{0.5in} m_{e} = m_{h} ;
\nonumber \\
B(p) &=& \frac{\alpha_{e}^{2} g}{16 \pi^{4}} \exp\left(
-\frac{\rho^{2}p^{2}}{16}\right) , \hspace{0.5in} m_{h} \gg m_{e};
\ \  \alpha_{h} \left( \frac{m_{h}}{m_{e}} \right)^{2} \ll
\alpha_{e}.
\end{eqnarray}
%-------------------------------------------------------------------------

Thus the CPA Green's function of the 2D indirect exciton is
determined by the solution of the self-consistent equations
Eqs.~(\ref{green_0}) and~(\ref{Q_def1}). In the weak-scattering
limit (~$g \ll e^{2}/\epsilon D M$~) we use the second-order Born
approximation for $Q$ similar to
Refs.~[\onlinecite{Gevorkyan,John_Stephen}], expanding $Q$
(Eq.~(\ref{Q_def1})) in a Taylor series to the first order in
$B(|\mathbf{p} - \mathbf{q}|)$ (which is the first order in $g$),
and we replace Eq.~(\ref{Q_def1}) by:
%-------------------------------------------------------------------------
\begin{eqnarray}
\label{Q_def} Q (\mathbf{p}, \omega) = \frac{\pi}{2} \int \delta
\left(\omega -  \frac{q^{2}}{2M}\right) B(|\mathbf{p} -
\mathbf{q}|) \frac{d^{d}q}{(2\pi)^{d}} .
\end{eqnarray}
%-------------------------------------------------------------------------
Substituting $B(p)$ from Eq.~(\ref{B_p}) into Eq.~(\ref{Q_def}),
we obtain for $Q (\mathbf{p}, \omega)$
%-------------------------------------------------------------------------
\begin{eqnarray}
\label{Q_res}  Q (\mathbf{p}, \omega) &=& \frac{(\alpha_{e} -
\alpha_{h})^{2}M g}{16\pi^{4}} \exp\left( -\frac{\rho^{2}}{8}
(p^{2} + 2M\omega)\right) J_{0}\left( \frac{\rho^{2}}{4}
\sqrt{2M\omega}p \right), \hspace{0.5in} m_{e} = m_{h} ;
\nonumber \\
   Q (\mathbf{p}, \omega) &=& \frac{\alpha_{e}^{2} M g}{64 \pi^{4}}  \exp\left(
-\frac{\rho^{2}}{16} (p^{2} + 2M\omega)\right) J_{0}\left(
\frac{\rho^{2}}{8} \sqrt{2M\omega}p \right), \nonumber \\
&& m_{h} \gg m_{e}; \ \ \alpha_{h} \left( \frac{m_{h}}{m_{e}}
\right)^{2} \ll \alpha_{e},
\end{eqnarray}
%-------------------------------------------------------------------------
where $J_{0}(z)$ is a Bessel function of the first kind. The
second-order Born Green's function of the single indirect exciton
in the
   random field, $G^{(0)}(\mathbf{p}, \omega)$, is derived by
substituting $Q (\mathbf{p}, \omega)$ from
   Eq.~(\ref{Q_res}) into Eq.~(\ref{green_0}).

%-------------------------------------------------------------------------
%-------------------------------------------------------------------------

\section{The Spectrum of Collective Excitations}
\label{collective_spectrum}

At small densities $n$ ($n \rho^2 \ll 1$), the system of indirect
excitons at low temperatures is a two-dimensional weakly
nonideal Bose
   gas with dipole moments  ${\bf d}$ normal to wells  ($d \sim eD$).
The distinction between excitons and bosons manifests itself in
exchange effects (see, e.g.,
Refs.[\onlinecite{Berman}],[\onlinecite{Berman_Willander}],
and~\cite{Halperin,Keldysh}). These effects are suppressed for
excitons with spatially separated $e$ and $h$ in a dilute system
($n\rho^2 \ll 1$)  at large $D$ ($D \gg a$),  because at large
$D$, the exchange interaction in the spatially separated system is
suppressed relative to the $e-h$ system in a single well due to
the smallness of the tunneling exponent  $T \sim
\exp[-(D/2a)^{1/4}]$ connected with the penetration of the
dipole-dipole interaction through the barrier between the two
wells\cite{Berman}. Hence, when $D \gg a$, exchange phenomena,
connected with the distinction between excitons and bosons, can be
neglected, and therefore, the system of indirect excitons in CQWs
can be treated by diagram techniques employed for boson
systems\cite{Abrikosov}. Two indirect excitons in a dilute system
interact as $U(R) = e^{2}D^{2}/(\epsilon R^{3})$, where $R$ is the
distance between exciton dipoles along quantum well planes.

As mentioned above, in the two-dimensional case the contribution
of the dipole interactions can be represented by the sum of the
ladder diagrams given in Fig.1. The Bogoliubov approximation for a
two-dimensional weakly interacting Bose gas is not valid due to
the divergence of the two-dimensional scattering amplitude in the
Born approximation\cite{Yudson}. Therefore, for $n \rho^2 \ll 1$,
we take into account the direct dipole-dipole repulsion between
excitons within the framework of two-dimensional Bose gas theory
in the ladder approximation\cite{Yudson}. Since the characteristic
frequencies and momenta which give the greatest contribution to
the single exciton Green's function $G^{(0)} (\mathbf{p}, \omega)$
in the ladder approximation are\cite{Yudson} $\omega \epsilon
D/e^{2} \sim n/\log[a^{2}/(nD^{4})] \ll 1$ and $p\rho(D) \sim M
\sqrt{n/\log[a^{2}/(nD^{4}))]} \ll 1$ (at $n\sqrt{aD^{3}} \ll 1$
and $D \gg a$), respectively,  we approximate $Q (\mathbf{p},
\omega)$ by $Q (\mathbf{p} = \mathbf{0}, \omega = 0)$ (see
Eq.~(\ref{Q_res}))
%-------------------------------------------------------------------------
\begin{eqnarray}
\label{Q_res_ap}  Q (\mathbf{p}, \omega) &=& Q = \frac{(\alpha_{e}
- \alpha_{h})^{2}M g}{16\pi^{4}} , \hspace{0.5in} m_{e} = m_{h} ;
\nonumber \\
   Q (\mathbf{p}, \omega) &=& Q = \frac{\alpha_{e}^{2} M g}{64 \pi^{4}}
, \hspace{0.5in}
   m_{h} \gg m_{e}; \ \ \alpha_{h} \left( \frac{m_{h}}{m_{e}}
\right)^{2} \ll \alpha_{e}.
\end{eqnarray}
%-------------------------------------------------------------------------
Note, that the replacing  $Q (\mathbf{p}, \omega)$ by constant $Q
(\mathbf{p} = \mathbf{0}, \omega = 0)$ does not constitute another
additional approximation, but  follows from Eq.~(\ref{Q_res}) at
small frequencies and momenta, which is applicable for the dilute
2D dipole gas as mentioned above. We show below that the constant
$Q$ actually does not affect the ladder approximation vertex
$\Gamma (\mathbf{p}, \mathbf{p}',0)$ for the weakly interacting
Bose gas\cite{Abrikosov}.

We pursue the 2D ladder approximation for the two-particle vertex
$\Gamma (p,p';P)$  (Fig.1) analogously to
Ref.~[\onlinecite{Yudson}]
%-------------------------------------------------------------------------
\begin{eqnarray}\label{Gamma_Green}
\Gamma (p,p';P) = U (\mathbf{p} - \mathbf{p}') + s\int_{}^{}
\frac{d^{3} q}{(2\pi )^3} U (\mathbf{p} - \mathbf{q})
G^{(0)}\left(\frac{P}{2} + q \right)G^{(0)} \left( \frac{P}{2} - q
\right)\Gamma (q,p';P),
\end{eqnarray}
%-------------------------------------------------------------------------
where the arguments of the vertex and the Green's function are 3D
momentum-frequency vectors (e.g., $p = \{\mathbf{p}, \omega \}$),
and $s$ is the level degeneracy (equal to 4 for excitons in GaAs
quantum wells). Using the single exciton Green's function $G^{(0)}
(\mathbf{p}, \omega)$ determined by Eq.~(\ref{green_0}) with $Q
(\mathbf{p}, \omega) = Q$ from Eq.~(\ref{Q_res_ap}), we obtain the
integral equation for $\Gamma$:
%-------------------------------------------------------------------------
\begin{eqnarray}\label{Gamma_Int}
\Gamma (\mathbf{p},\mathbf{p}';P) &=& U (\mathbf{p} - \mathbf{p}')
+ s\int_{}^{} \frac{d^{2} q}{(2\pi )^2} \frac{U (\mathbf{p} -
\mathbf{q}) \Gamma (\mathbf{q},\mathbf{p}';P)}{\frac{\kappa ^2}{M}
+\Omega -
\frac{\mathbf{P}^{2}}{4M} - \frac{q^2}{M} + 2 i Q} \nonumber\\
\mu &=& \frac{\kappa ^2}{2M} = n\Gamma _{0} = n \Gamma (0,0;0),
\end{eqnarray}
%-------------------------------------------------------------------------
where $P = \{\mathbf{P},\Omega \}$, and $\mu $ is the chemical
potential of the system.

The integral equation Eq. (\ref{Gamma_Int}) for the vertex  can be
solved analytically in the approximation  $\kappa \ll
\sqrt{n}$.\cite{Yudson} This inequality must be fulfilled
simultaneously with the condition of low density $n\rho^{2}(D) \ll
1$ (at $n\sqrt{aD^{3}} \ll 1$ and $D \gg a$) which is necessary
for the applicability of the ladder approximation.  The solution
of the integral equation for the vertex $\Gamma $ of this system
can be expressed through the solution of the equation for the
scattering amplitude  $f_0(\kappa ) \equiv f(\mathbf{\kappa},
\mathbf{\kappa})$ of an isolated pair of interacting particles  in
a two-dimensional system  with the repulsive potential $U(R) = e^2
D^2/(\epsilon R^3)$:
%-------------------------------------------------------------------------
\begin{equation}\label{f_0}
f_0 (\kappa) = \frac{\left( \frac{\pi i}{2 \kappa } \right)
^{1/2}} {\log \left(\frac{\kappa M e^2 D^2}{\epsilon} \right)} .
\end{equation}
%-------------------------------------------------------------------------
The relation of the vertex $\Gamma (\mathbf{p}, \mathbf{p}',0)$
(Fig.1.), which represents the effective interaction, to the
two-dimensional scattering amplitude $f_{0}(\mathbf{p}',
\mathbf{p})$ is
%-------------------------------------------------------------------------
\begin{eqnarray}
\label{ver_amp} \Gamma (\mathbf{p},\mathbf{p}';0) &=& \left[ - f_0
(\mathbf{p}',\mathbf{p}) \frac{2}{M}\left(\frac{2\pi p'}{i}
\right)^{1/2}\right] + s \int_{}^{} \frac{d^{2} K}{(2\pi )^2}
\left[ - f_0 (\mathbf{K},\mathbf{p}) \frac{2}{M}\left(\frac{2\pi
K}{i} \right)^{1/2}  \right]  \Gamma (\mathbf{K},\mathbf{p}';0)
\nonumber \\ &\times& \left\{\frac{1}{\frac{\kappa ^2}{M} -
\frac{K^2}{M} + 2 i Q}  - \frac{1}{\frac{p'^2}{M}   -
\frac{K^2}{M} + i\delta} \right\} ,
\end{eqnarray}
%-------------------------------------------------------------------------
where $\delta \rightarrow 0$. In the first order in the scattering
amplitude $f_{0}$ of the weakly interacting Bose-gas we neglect
the second term in the r.h.s. of Eq.~(\ref{ver_amp}). Therefore,
the ladder approximation vertex $\Gamma$ does not depend on the
constant $Q$ for the weakly interacting Bose gas to first order in
the scattering amplitude. Hence, the vertex of the dipole
repulsion for the rare exciton gas in the random field will be the
same as for the clean system without random field.

   Here the characteristic  momentum
$\kappa $, unlike in the three-dimensional system, is not equal to
zero but rather is determined from the relation\cite{Yudson}
%-------------------------------------------------------------------------
\begin{equation}\label{Kappa}
\kappa ^2 = -4nf_0 (\kappa) \left(\frac{2\pi \kappa }{i}
\right)^{1/2} .
\end{equation}
%-------------------------------------------------------------------------
This is a specific feature of two-dimensional Bose system
connected with the logarithmic divergence of the two-dimensional
scattering amplitude at zero energy. A simple analytical solution
for the chemical potential can be obtained if $\kappa M e^{2}
D^2/\epsilon \ll 1$.  In this limit, the chemical potential $\mu $
takes the form:
%-------------------------------------------------------------------------
\begin{equation}\label{Mu}
\mu =  \frac{\kappa ^2 }{2M} = \frac{8\pi n}{2M \log \left(
\frac{\epsilon^{2}}{8\pi s^2 n M^2 e^4 D^4} \right)} .
\end{equation}
%-------------------------------------------------------------------------

Since the only difference between the Green's function of the
isolated exciton with and without the random field is the term $i
Q$ in Eq.~(\ref{green_0}), all internal blocks of the ladder
approximation diagrams for the system in the random field give the
same self-energy as without the random field. This can be shown by
repeating the procedure of the derivation of the 2D ladder
approximation self-energy for the ``clean'' system\cite{Yudson} by
using integration measure $d(\omega + iQ)$ instead of $d\omega$.
The difference in the Green's function will appear only in the
external lines as the replacing of $d\omega$ by $d(\omega + iQ)$.
Introducing the Green's function of the Bose condensate and normal
and anomalous Green's functions of the noncondensate analogously
to Ref.~[\onlinecite{Berman_Ruvinsky}],  we use the finite
temperature Green's functions, replacing $\omega$ in the zero
temperature Green functions
   by $i \omega_{k}$, where $\omega _{k} = 2\pi kT$ ($k$ is
an integer\cite{Abrikosov}; we set the Boltzmann constant $k_{B} =
1$). This procedure is valid because at small temperatures ($T \ll
\mu$) we assume the collective spectrum to be the same as the
zero-temperature spectrum\cite{Huang,Huang_Meng,Lopatin}. For the
2D system the temperatures where  superfluidity exists can be
assumed small, because they are required to be below the
Kosterlitz-Thouless temperature $T_{c}$\cite{Kosterlitz} (we show
below that $T_{c}<\mu $).

We therefore have the condensate Green's function
$D(\mathbf{p}, i\omega_{k})$
%-------------------------------------------------------------------------
\begin{eqnarray}
\label{d0} D^{(0)}({\bf p},i\omega _{k}) = \frac{- i(2\pi )^{2}
n_{0}\delta ({\bf p})}{i\omega _{k} + \mu + i Q},
\end{eqnarray}
%-------------------------------------------------------------------------
where $n_{0}$ is the density of Bose condensate. Since at small
temperatures $(n - n_{0})/n \ll 1$, according to the ladder
approximation\cite{Abrikosov} we use $n$ below  instead of
$n_{0}$. $G(\mathbf{p}, i\omega_{k})$ and $F(\mathbf{p},
i\omega_{k})$ are the normal and anomalous Green functions of the
overcondensate:
%-------------------------------------------------------------------------
\begin{eqnarray}
\label{g0} G({\bf p},i\omega _{k}) &=& - \frac{i \omega _{k} +
\varepsilon _{0}(p)
   + \mu + i Q}{
\omega _{k}^{2} + \varepsilon ^{2}(p) - 2i (\mu - \varepsilon
_{0}(p))Q}  ; \nonumber\\
   F({\bf
p},i\omega _{k}) &=&  - \frac{\mu }{ \omega _{k}^{2} + \varepsilon
^{2}(p) - 2i(\mu - \varepsilon _{0}(p)) Q} ,
\end{eqnarray}
%-------------------------------------------------------------------------
where $\varepsilon _{0}(p)$ is the spectrum of noninteracting
excitons; the spectrum of interacting excitons  has the form
$\varepsilon (p) = \sqrt{\left(p^{2}/(2M) + \sqrt{\mu^{2} -
Q^{2}}\right)^{2} - (\mu^{2} - Q^{2})}$, and for small momenta $p
\ll \mu$ the excitation spectrum is acoustic $\varepsilon (p) =
c_{s} p$, where $c_{s} = \sqrt{\sqrt{\mu^{2} - Q^{2}}/M}$ is the
velocity of sound.

  From the expression for the spectrum $\varepsilon (p)$ we see
that for $Q > \mu$ the Bose condensate state becomes unstable,
because the spectrum $\varepsilon (p)$ becomes imaginary. So, for
$Q > \mu$ the random field destroys superfluidity, even at $T =
0$. This condition for the instability of the condensate itself is
very approximate, because we have used the ladder approximation,
which is valid only if almost all paricles are in the condensate
($(n - n_{0})/n \ll 1$), which is not the case for large random
field contribution $Q$. Since at $Q < \mu$ the spectrum of the
system is acoustic and satisfies the Landau criterium of
superfluidity, the system becomes a two-component liquid,
consisting of the superfluid and normal component in the presence
of the random field even at $T = 0$\cite{Abrikosov}. As $Q$ grows,
the system undergoes a transition to the exciton glass state (for
the a lattice model of bosons, this transition was considered in
Ref.[\onlinecite{Fisher}]).

%-------------------------------------------------------------------------
%-------------------------------------------------------------------------

\section{The Kosterlitz-Thouless Phase Transition}
\label{KT_f_tr}

The density of the superfluid component $n_{s}(T)$ can be obtained
using the relation $n_{s}(T) = n - n_{n}(T)$, where $n_{n}(T)$ is
the density of the normal component.

   The density of the normal component $n_{n}(T)$, which is dissipated at
the walls and impurities, can be calculated using the Kubo formula
as the response of the total momentum to an external
velocity\cite{Mahan}:
%-------------------------------------------------------------------------
\begin{eqnarray}
\label{gr2} n_{n}  =  - \lim_{\omega \to 0} \left[ \frac{Im(\Pi
(i\omega ))}{i \omega } \right] ,
\end{eqnarray}
%-------------------------------------------------------------------------
where $\Pi (i\omega )$ is the polarization operator with zero
transferred momentum
%-------------------------------------------------------------------------
\begin{eqnarray}
\label{pi} \Pi (i\omega ) =  \frac{1}{2M} s\sum_{{\bf p}}^{} p^{2}
T \sum_{\omega '_{k}}^{} {\cal F}({\bf p}, i \omega '_{k} + i
\omega) {\cal F}({\bf p}, i \omega '_{k})    ,
\end{eqnarray}
%-------------------------------------------------------------------------
and ${\cal F}({\bf p}, i \omega_{k})$  is the total
single-particle Matsubara Green's function of an indirect exciton
%-------------------------------------------------------------------------
\begin{eqnarray} \label{F} {\cal F}({\bf p}, i \omega '_{k}) =
D({\bf p}, i \omega '_{k}) + G({\bf p}, i \omega '_{k}) .
\end{eqnarray}
%-------------------------------------------------------------------------
The renormalization of the vertex by the interaction is neglected
in the polarization operator Eq.~(\ref{pi}). When the interaction
is taken into account in the ladder approximation, a term which is
small with respect to the parameter $M\Gamma \ll 1$ appears
($\Gamma$ is the vertex in the ladder approximation). For a
two-dimensional rarefied system of indirect excitons this
parameter has the form $4\pi/\log\left[(\epsilon^{2}/(8\pi s^2 n
M^{2}e^{4}D^{4})\right] \ll 1 $.

We now substitute the Green's functions of the condensate
Eq.~(\ref{d0}) and noncondensate Eq.~(\ref{g0}) particles into
Eq.~(\ref{F}). Next, substituting the expression Eq.~(\ref{F})
into Eqs. (\ref{pi}) and (\ref{gr2}) we have
%-------------------------------------------------------------------------
\begin{eqnarray}
\label{nn4} n_{n} = n_{n}^{0} + s\frac{N}{M} \int_{}^{}\frac{d{\bf
p}}{(2\pi )^{2}} p^{2}\mu \frac{\varepsilon _{0}(p)}{ \varepsilon
^{4}(p)}Q       .
\end{eqnarray}
%-------------------------------------------------------------------------
Here $N$ is the total number of particles, and $n_{n}^{0}$ is the
density of the normal component in a pure system with no
impurities:
%-------------------------------------------------------------------------
\begin{eqnarray}
\label{Bose} n_{n}^{0}  = - s\frac{1}{2M} \int_{}^{} \frac{d{\bf
p}}{(2\pi )^{2}} p^{2} \frac{\partial n_{0}(p)}{\partial
\varepsilon}.
\end{eqnarray}
%-------------------------------------------------------------------------
where $n_{0}(p)=(e^{\varepsilon (p)/T} - 1)^{-1}$  is the
distribution of an ideal Bose gas of temperature excitations.

The first term in Eq.~(\ref{nn4}), which does not depend on $Q$, is the
contribution to the normal
component due to scattering of quasiparticles with an acoustic
spectrum in an ordered system at $T \neq 0$. In a two-dimensional system,
%-------------------------------------------------------------------------
\begin{eqnarray}
\label{nn00} n_{n}^{0} = s \frac{3 \zeta (3) }{2 \pi }
\frac{T^3}{c_{s}^{4}(n,Q) M},
\end{eqnarray}
%-------------------------------------------------------------------------
where $\zeta (z)$ is the Riemann zeta function ($\zeta (3) \simeq
1.202$).    The second term in Eq.~(\ref{nn4}) is the contribution
to the normal component due to the interaction of the particles
(excitons) with the random field,
%-------------------------------------------------------------------------
\begin{eqnarray}
\label{nn55} n_{n} = n_{n}^{0} + s\frac{n Q}{2Mc_{s}^{2}(n,Q)} =
s\frac{3 \zeta (3) }{2 \pi } \frac{T^3}{c_{s}^{4}(n,Q) M} + s\frac{n
Q}{2Mc_{s}^{2}(n,Q)}.
\end{eqnarray}
%-------------------------------------------------------------------------
The density of the superfluid component is $n_{s} = n - n_{n}$.
  From Eqs.~(\ref{nn00}) and~(\ref{nn55}) we can see that the
random field decreases the density of the superfluid component.

In a 2D system, superfluidity appears below the
Kosterlitz-Thouless transition temperature $T_{c} = \pi
n_{s}/(2M)$,\cite{Kosterlitz} where only coupled vortices are
present. Using the expressions (\ref{nn00}) and~(\ref{nn55}) for
the density $n_{s}$ of the superfluid component, we obtain an
equation for the Kosterlitz-Thouless transition temperature
$T_{c}$. Its solution is
%-------------------------------------------------------------------------
\begin{eqnarray}
\label{tct} T_c = \left[\left( 1 +
\sqrt{\frac{32}{27}\left(\frac{M T_{c}^{0}}{\pi n'}\right)^{3} +
1} \right)^{1/3}   - \left( \sqrt{\frac{32}{27} \left(\frac{M
T_{c}^{0}}{\pi n'}\right)^{3} + 1} - 1 \right)^{1/3}\right]
\frac{T_{c}^{0}}{ 2^{1/3}}   .
\end{eqnarray}
%-------------------------------------------------------------------------
Here $T_{c}^{0}$ is an auxiliary quantity, equal to the
temperature at which the superfluid density vanishes in the
mean-field approximation (i.e., $n_{s}(T_{c}^{0}) = 0$),
%-------------------------------------------------------------------------
\begin{equation}
\label{tct0} T_c^0 = \left( \frac{2 \pi n' c_s^4 M}{3 \zeta (3)}
\right)^{1/3}  .
\end{equation}
%-------------------------------------------------------------------------
In Eqs.~(\ref{tct}) and (\ref{tct0}), $n'$ is
%-------------------------------------------------------------------------
\begin{eqnarray}
\label{nexx} n' = n  - s\frac{n Q}{2Mc_{s}^{2}} .
\end{eqnarray}
%-------------------------------------------------------------------------

%-------------------------------------------------------------------------
%-------------------------------------------------------------------------

\section{Discussion}
\label{discussion}

The dependence of the Kosterlitz-Thouless transition temperature
$T_{c}$ as a function of the total exciton density $n$ for
different $Q$, obtained from Eq.~(\ref{tct}), is presented in
Fig.~2. It can be seen in Fig. 2 that the random field decreases
the Kosterlitz-Thouless transition temperature. This trend was
pointed out before, for weak coupling with the random field, in
Ref. [\onlinecite{Berman_Ruvinsky}]. Figs. 3 and 4 show the
dependence of the Kosterlitz-Thouless transition temperature on
the random field parameter $Q$ and the spatial separation between
the electrons and holes.

The results of the approximation used in Ref.
[\onlinecite{Berman_Ruvinsky}] can be obtained from the normal
density $n_{n}$, derived in the present work (Eq.~(\ref{nn55})),
as a first order in the expansion of $n_{n}$ in Eq.~(\ref{nn55})
respect to $Q/\mu$, which corresponds to the case when the random
field is weaker than the dipole-dipole repulsion and $Q/\mu \ll
1$. (Operationally, this corresponds to replacing the speed of
sound by its value at $Q = 0$, i.e., $c_{s} \rightarrow
\sqrt{\mu/M}$, in Eq.~(\ref{nn55}).) For realistic experimental
parameters, the random field is not always smaller than the
dipole-dipole repulsion. (E.g., in Ref. [\onlinecite{Snoke-apl}],
the luminescence linewdith due to inhomogeneous broadening at low
density is approximately 2 meV; more recent GaAs
structures\cite{Snoke_paper} have inhomogeneously broadened
linewidths closer to 1 meV.) Figs.~5 and~6 show that the approach
used in the present work results in the Kosterlitz-Thouless
temperature being smaller than the transition temperature obtained
from the approximation used in Ref.
[\onlinecite{Berman_Ruvinsky}], which is denoted here by $PT$ (for
perturbation theory; see below). The difference in the
Kosterlitz-Thouless temperature between these two approaches
increases when random field $Q$ increases and exciton density $n$
decreases (Figs.~5 and~6). The results of this comparison are
reasonable, because the approximation used in Ref.
[\onlinecite{Berman_Ruvinsky}] implies a first-order perturbation
theory with respect to $Q/\mu$ for the Green's function, and the
inequality $Q/\mu \ll 1$ holds for small $Q$ and not very small
densities $n$ ($\mu$ increases as $n$ increases). Note that in the
present work the parameter $Q/\mu$ is not required to be small.

This work shows that although the random field depletes the
condensate, Kosterlitz-Thouless superfluidity is still possible in
a system of spatially indirect excitons. However, for realistic
experimental parameters, the possibility of superfluidity is
marginal. For the structure parameters used in the plots of Figs.
2-4, with effective $D=15 \ nm$, the binding energy of the
indirect excitons at high electric field is 3.5 $meV$ and the
excitonic Bohr radius is around 150 \AA. This implies that $na^2
\cong 1$, the point at which the exciton gas becomes plasma-like,
at a density of around $5\times 10^{11}$ cm$^{-2}$. If the exciton
density is to be well below this value, say an order of magnitude,
then as seen in Fig. 2, the disorder parameter $Q$ must be around
0.1-0.2 meV. By comparison, the
  half-width, half maximum of the exciton luminescence line at low
temperature and low
density in recent samples, due to inhomogeneous broadening, is around 0.5 meV
\cite{Snoke_paper}. It is not clear exactly how the inhomogeneous
line width and the disorder
factor $Q$ are related. The line width is likely sensitive to
disorder with length scales short
compared to the excitonic Bohr radius, while the $Q$ parameter used
here is assumed to be a measure
of disorder on length scales long compared to the excitonic Bohr
radius. Nevertheless, there is
still a problem with Kosterlitz-Thouless superfluidity at low
density, no matter what the
temperature.

The story may change when the excitons are trapped in a confining
potential, in which case true Bose condensation is possible in the
Stringari-Pitaevskii limit.\cite{Daflovo} However, these results
indicate that the role of disorder must be taken into account for
any realistic treatment of excitonic condensation.

%-------------------------------------------------------------------------

\section*{Acknowledgements}
O. L.~B. wishes to thank Prof.~Dan~Boyanovsky for many useful and
stimulating discussions. Yu. E.~L. was supported by the INTAS
grant. D. W.~S. and R. D.~C. have been supported by the National
Science Foundation.

%-------------------------------------------------------------------------
%-------------------------------------------------------------------------

\newpage

%\begin{large}

\begin{center}
{\bf Captures to Figures (1-6)}
\end{center}

Fig.1. Diagrammatic representation of the equation for the vertex
$\Gamma $ in the momenta-frequency representation $({\bf P},
\Omega)$.

Fig.2.  Dependence of temperature of Kosterlitz-Thouless
transition $T_{c} = T_{c}(n)$ (in units of $K$; for GaAs/AlGaAs:
$M = 0.24 m_{0}$, where $m_{0}$ is the free electron mass;
$\epsilon = 13$) on the exciton density $n$ (in units of
$cm^{-2}$) at the interwell distance $D = 15 \ nm$, for different
random fields $Q$ (in units of $meV$): $Q = 0$ -- solid curve
(straight line); $Q = 0.1 \ meV$ -- dotted curve; $Q = 0.2 \  meV$
-- dashed curve; $Q = 0.3 \ meV$ --   long-dashed  curve; $Q = 0.4
\ meV$ -- dashed-dotted  curve; $Q = 0.5 \ meV$ -- solid curve
(bottom right hand corner).

Fig.3.  Dependence of temperature of Kosterlitz-Thouless
transition $T_c = T_c (Q)$ (in units of $K$; for GaAs/AlGaAs: $M =
0.24 m_{0}$; $\epsilon = 13$) on the random field $Q$ (in units of
$meV$) at the interwell distance $D = 15 \ nm$, for different
exciton densities $n$: $n = 5\times 10^{10} \ cm^{-2}$ -- solid
curve; $n = 1.0\times 10^{11} \ cm^{-2}$ -- dotted curve; $n =
3.0\times 10^{11} \ cm^{-2}$  -- dashed curve.

Fig.4. Dependence of temperature of Kosterlitz-Thouless transition
$T_c = T_c (D)$ (in units of $K$; for GaAs/AlGaAs: $M = 0.24
m_{0}$; $\epsilon = 13$) on the interwell distance $D$ (in units
of $nm$) at the exciton density $n = 1\times 10^{11} \ cm^{-2}$,
for different random fields $Q$: $Q = 0$ -- solid curve; $Q = 0.3
\  meV$ -- dotted curve; $Q = 0.5 \ meV$ -- dashed curve; $Q = 0.6
\ meV$ -- dashed-dotted curve.

Fig.5.  Temperature dependence of Kosterlitz-Thouless transition
$T_{c} = T_{c}(n)$  based on CPA and the perturbation theory
 (PT) approximation used in Ref. \protect[\onlinecite{Berman_Ruvinsky}]
(in units of $K$; for GaAs/AlGaAs: $M = 0.24 m_{0}$; $\epsilon =
13$; $m_{0}$ is a mass of electron) on the exciton density $n$ (in
units $cm^{-2}$) at the interwell distance $D = 15 \ nm$, for
different random fields $Q$ (in units of  $meV$): $Q = 0.2 \ meV$
 -- solid curve, full CPA {\it vs.} dotted curve, PT; $Q = 0.3 \ meV$  --
dashed curve, full CPA {\it vs.} long-dashed curve, PT.

Fig.6.   Temperature dependence of Kosterlitz-Thouless transition
$T_{c} = T_{c}(Q)$  based on CPA and the perturbation theory
 (PT) approximation used in Ref. \protect[\onlinecite{Berman_Ruvinsky}]
(in units of $K$; for GaAs/AlGaAs: $M = 0.24 m_{0}$; $\epsilon
=13$) on the random field $Q$ (in units of $meV$) at the interwell
distance $D = 15 \ nm$, for different exciton densities $n$: $n =
5 \times 10^{10} \ cm^{-2}$ -- solid curve, full CPA {\it vs.}
dotted curve, PT; $n = 1 \times 10^{11} \ cm^{-2}$  -- dashed
curve, full CPA {\it vs.} long-dashed curve, PT.

\newpage

\begin{figure}
\includegraphics[width = 15cm, height = 7cm]{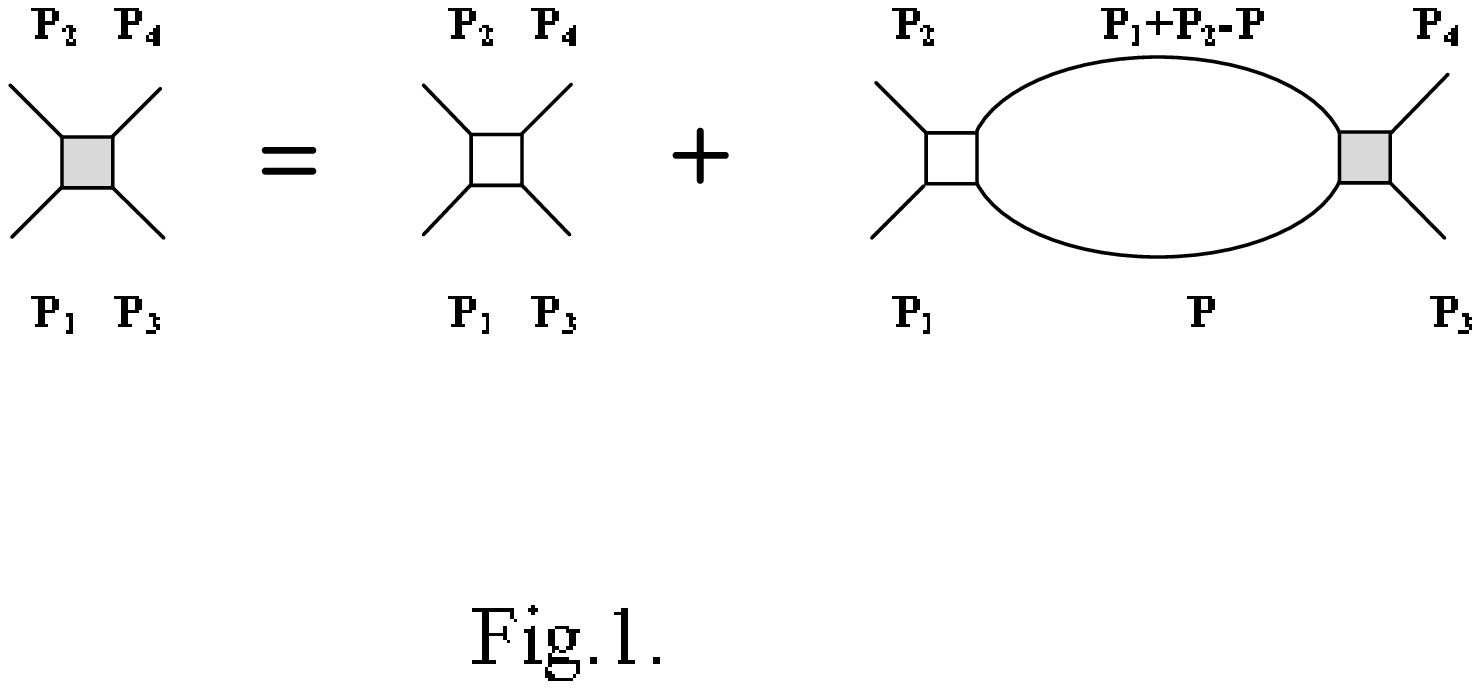}
%\caption{}
\end{figure}
\vspace{10cm}
\newpage
\newpage
\pagebreak

\begin{figure}
\rotatebox{270}{
\includegraphics[width = 17cm, height = 16cm]{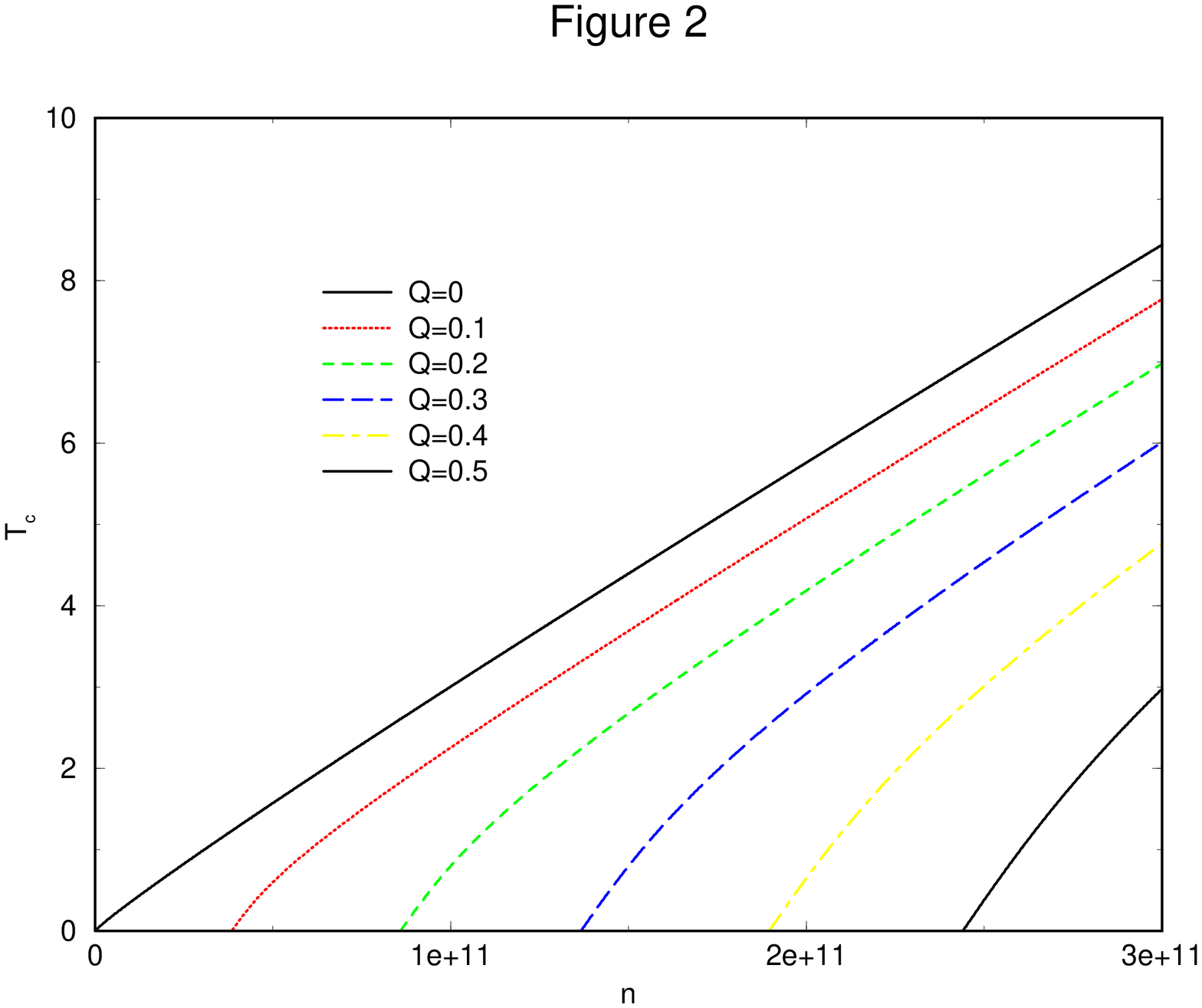}}
%\caption{}
\end{figure}

\newpage
\newpage

\begin{figure}
\rotatebox{270}{
\includegraphics[width = 17cm, height = 16cm]{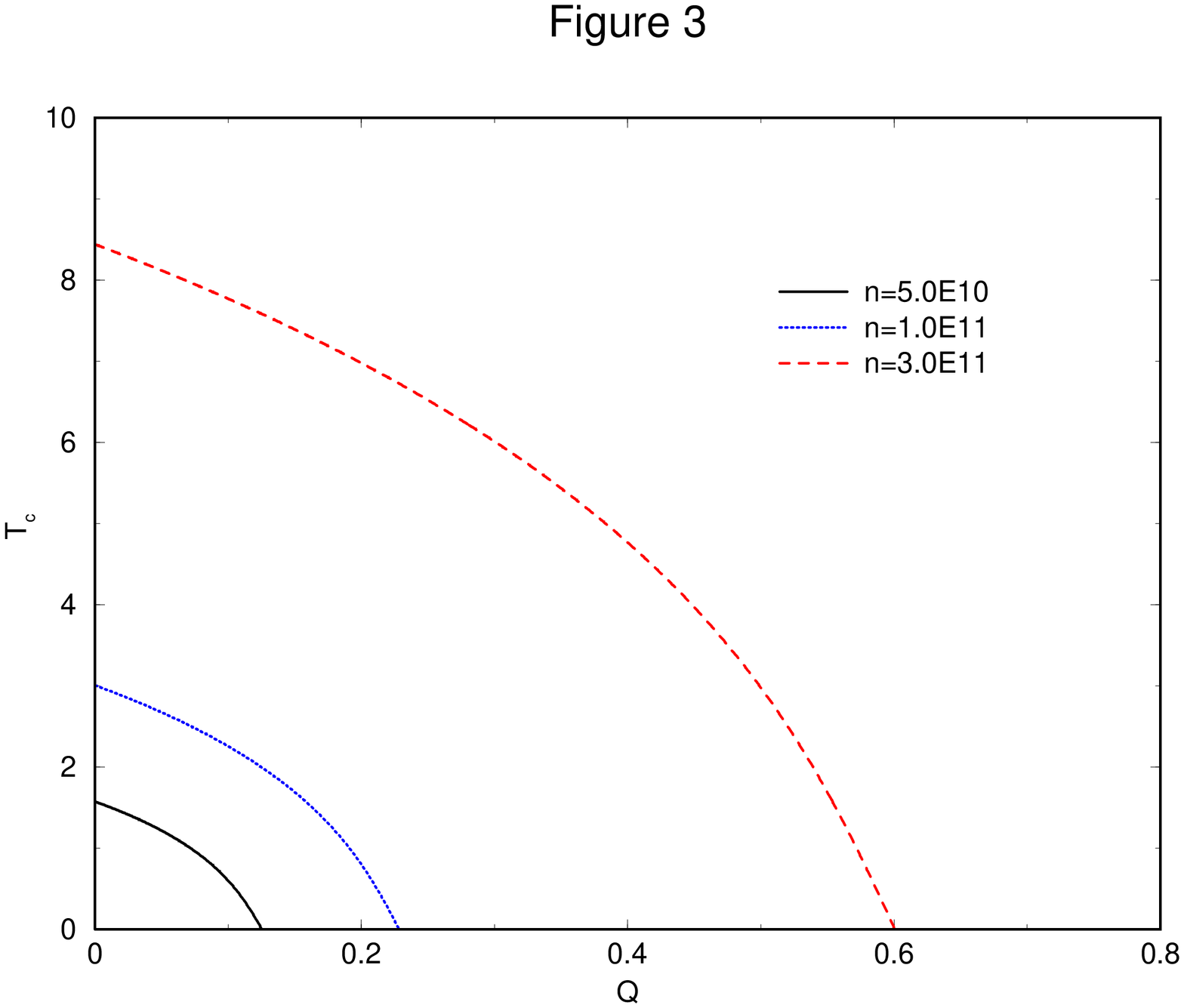}}
%\caption{}
\end{figure}

\newpage
\newpage

\begin{figure}
\rotatebox{270}{
\includegraphics[width = 17cm, height = 16cm]{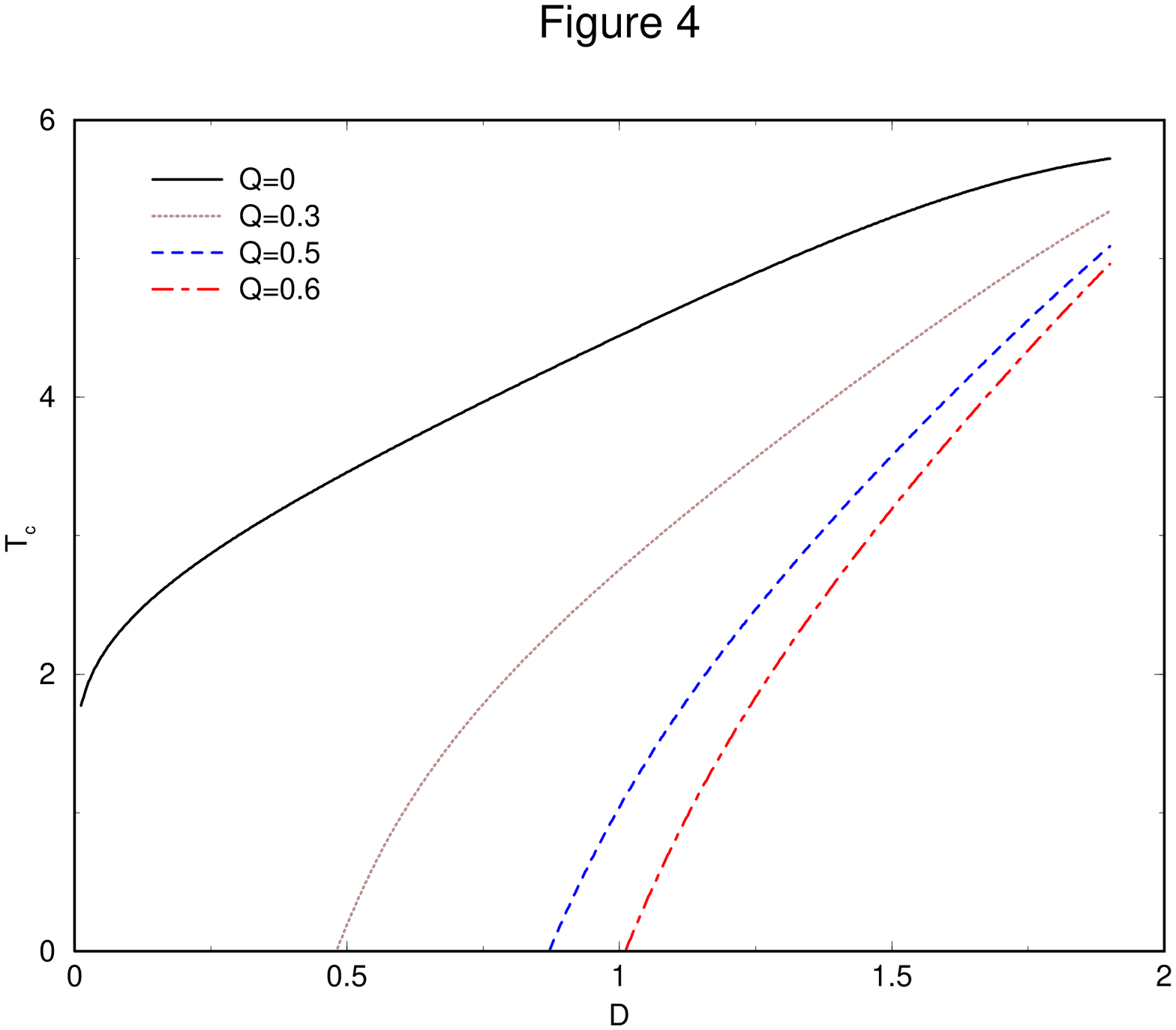}}
%\caption{}
\end{figure}

\newpage
\newpage

\begin{figure}
\rotatebox{270}{
\includegraphics[width = 17cm, height = 16cm]{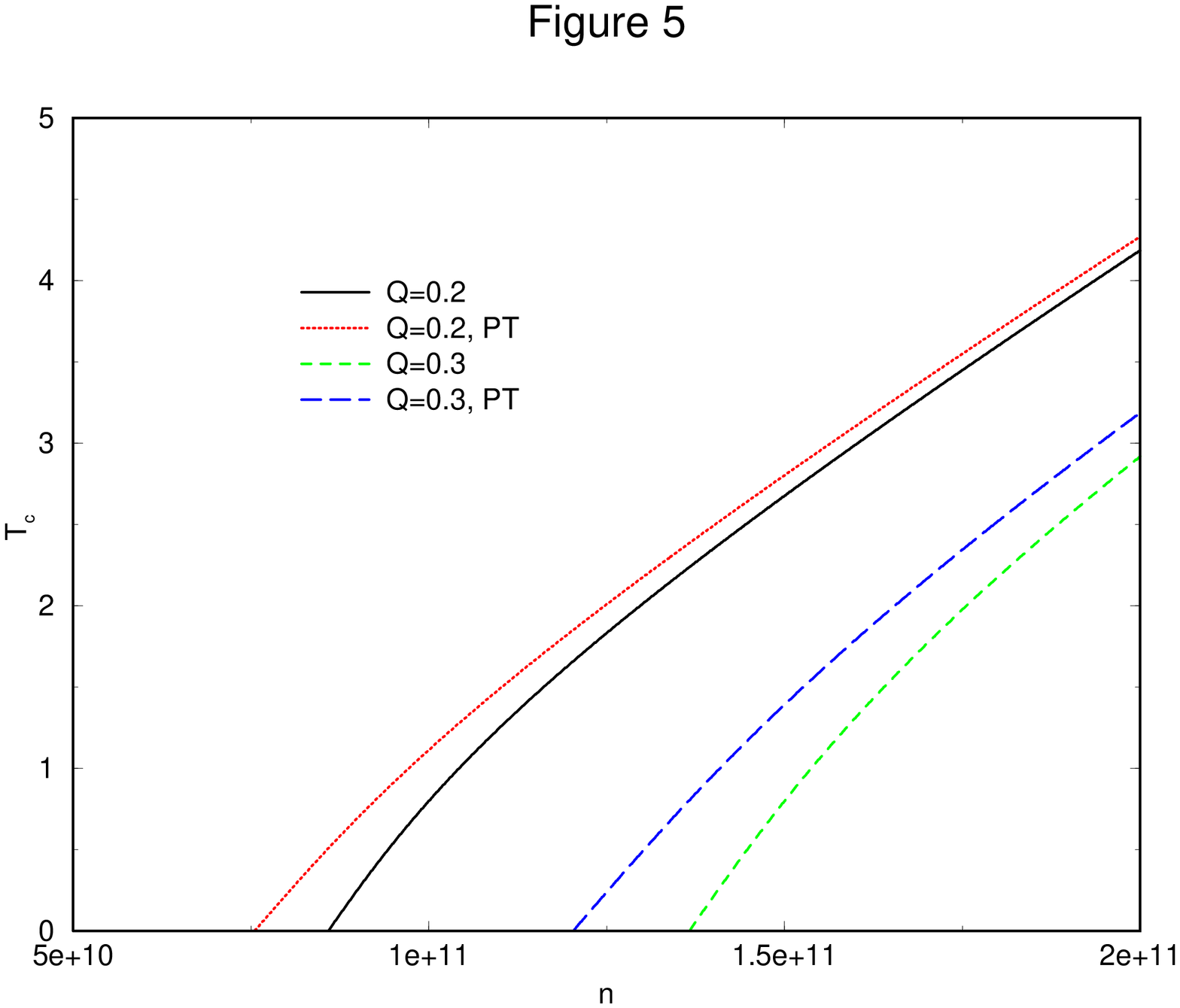}}
%\caption{}
\end{figure}

\newpage
\newpage

\begin{figure}
\rotatebox{270}{
\includegraphics[width = 17cm, height = 16cm]{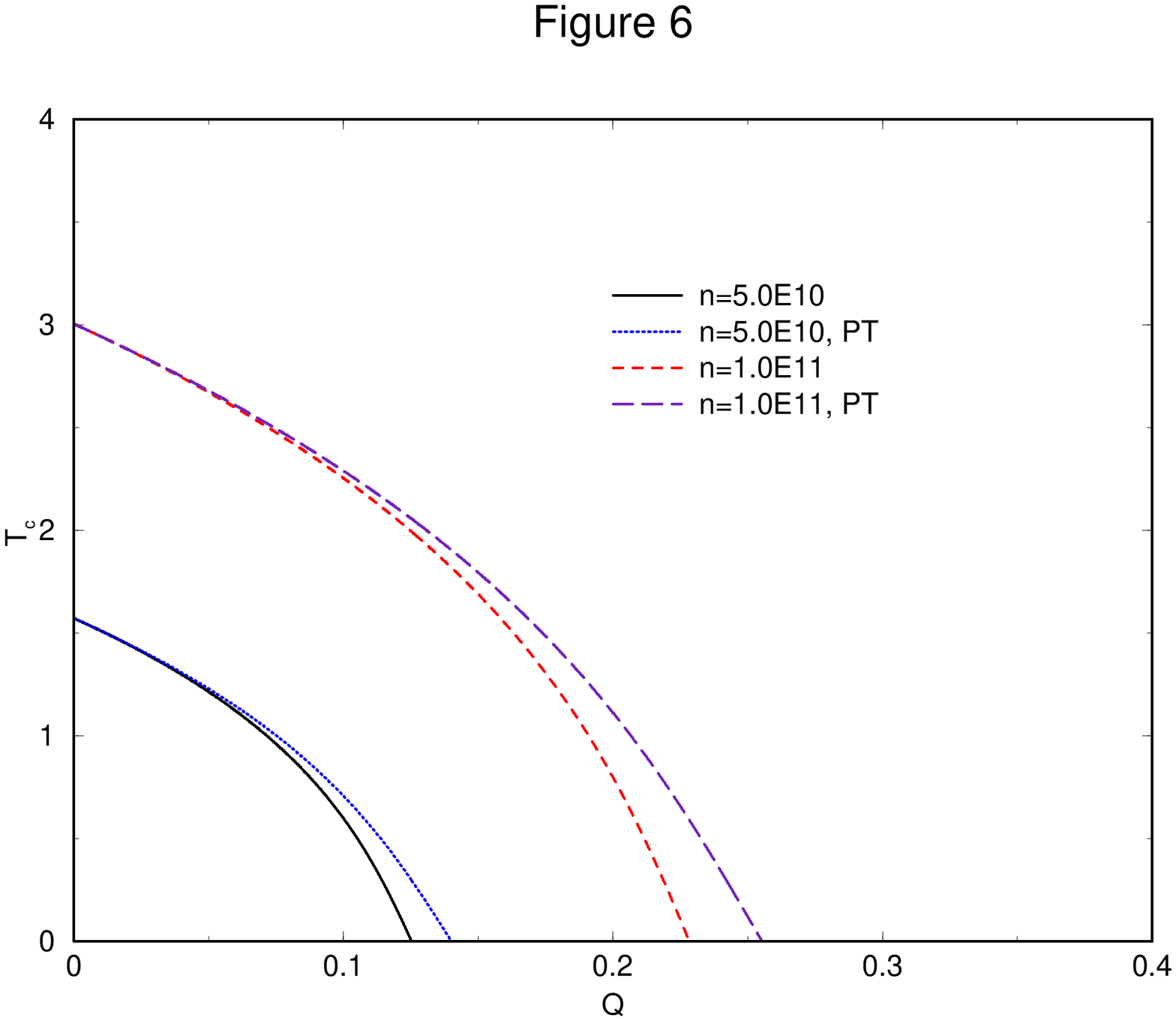}}
%\caption{}
\end{figure}


\begin{thebibliography}{99}

\bibitem{Lozovik} Yu.~E. Lozovik and V.~I. Yudson,
   Pis'ma ZhETF {\bf 22}, 26 (1975)
[JETP Lett. {\bf 22}, 26(1975)]; ZhETF {\bf 71}, 738 (1976) [JETP
{\bf 44}, 389 (1976)]; Sol. St. Comms. {\bf 18}, 628 (1976); Sol.
St. Comms. {\bf 18}, 628 (1976); Sol. St. Comms. {\bf 21}, 211
(1977).

\bibitem{Klyuchnik} A.~V. Klyuchnik and Yu.~E. Lozovik,  ZhETF {\bf
76}, 670 (1979) [JETP {\bf 49}, 335 (1979)]; Yu.~E. Lozovik and
A.~V. Poushnov, Phys.~Lett.~{\bf A 228}, 399 (1997).

\bibitem{Shevchenko} S.~I. Shevchenko, Phys.~Rev.~Lett. {\bf 72}, 3242
(1994).

\bibitem{Lerner} I.~V. Lerner and  Yu.~E. Lozovik, ZhETF {\bf 74}, 274 (1978)
[JETP {\bf 47}, 146 (1978)]; ZhETF {\bf 76}, 1136 (1979) [JETP
{\bf 49}, 576 (1979)]; ZhETF {\bf 78}, 1167 (1980) [JETP {\bf 51},
588 (1980)]; ZhETF {\bf 80}, 1488 (1981) [JETP {\bf 53}, 763
(1981)];

\bibitem{Dzjubenko} A.~B. Dzjubenko and Yu.~E. Lozovik, Fiz. Tverd. Tela
   {\bf 25}, 1519 (1983) [Solid State Phys.  {\bf 25}, 874 (1983) ];
Fiz. Tverd. Tela {\bf 26}, 1540 (1984) [Solid State Phys.  {\bf
26}, 938 (1984) ]; J. Phys. {\bf A 24}, 415 (1991); Yu.~E.
Lozovik, I.~V. Ovchinnikov, S.~Yu. Volkov, L.~V. Butov, and D.~S.
Chemla, Phys. Rev. {\bf B 65}, 235304 (2002).


\bibitem{Kallin}  C. Kallin and B.~I. Halperin,
Phys. Rev. {\bf B 30}, 5655 (1984); Phys. Rev. {\bf B 31}, 3635
(1985).


\bibitem{Knox} D.~S.~Chemla,~J.~B.~Stark, and W.~H.~Knox,
   In {\it "Ultrafast Phenomena VIII"},
Eds.~J.~-L. Martin et al.,~Springer, 21 (1993); C.~W. Lai, J.
Zoch, A.~C. Gossard, and D.~S. Chemla, Science {\bf 303}, 503
(2004).


\bibitem{Yoshioka} D. Yoshioka and A.~H. MacDonald,
J. Phys.S oc. of Japan, {\bf 59}, 4211 (1990).

\bibitem{Birman} J. Zang, D. Schmeltzer and J.~L. Birman, Phys. Rev.
Lett. {\bf 71}, 773
(1993).


\bibitem{Littlewood} Xu. Zhu, P.~B. Littlewood, M.~S. Hybertsen and T.~M. Rice,
   Phys. Rev. Lett. {\bf 74}, 1633 (1995).

\bibitem{Vignale}  S. Conti, G. Vignale and A.~H. MacDonald, Phys.
Rev.  {\bf B 57}, R6846 (1998).


\bibitem{Berman}  Yu.~E. Lozovik and O.~L. Berman, Pis'ma ZhETF {\bf 64}, 526
(1996) [JETP Lett. {\bf 64}, 573 (1996)]; ZhETF {\bf 111}, 1879
(1997) [JETP {\bf 84}, 1027 (1997)].



\bibitem{Berman_Tsvetus} Yu.~E. Lozovik, O.~L. Berman and
   V.~G. Tsvetus,  Phys. Rev.  {\bf B 59}, 5627 (1999).

\bibitem{Berman_Willander} Yu.~E. Lozovik, O.~L. Berman, and M.
Willander, J. Phys.: Condens. Matter {\bf 14}, 12457 (2002).


\bibitem{Ulloa} M.~A. Olivares-Robles and S.~E. Ulloa, Phys. Rev.
{\bf B 64}, 115302 (2001).



\bibitem{Snoke_paper} D. Snoke,
S. Denev, Y. Liu, L. Pfeiffer and K. West, Nature {\bf 418}, 754
(2002).



\bibitem{Snoke_paper_Sc} D. Snoke, Science {\bf 298}, 1368 (2002).


\bibitem{Chemla}    L.~V. Butov, A. Zrenner, G. Abstreiter,
G. Bohm and G. Weimann, Phys. Rev. Lett. {\bf 73}, 304 (1994);
L.~V. Butov, C.~W. Lai, A.~L. Ivanov, A.~C. Gossard, and D.~S.
Chemla, Nature {\bf 417}, 47 (2002); L.~V. Butov, A.~C. Gossard,
and D.~S. Chemla, Nature {\bf 418}, 751 (2002)).

\bibitem{Krivolapchuk} V.~V.Krivolapchuk, E.~S.Moskalenko, and A.~L. Zhmodikov,
Phys.~Rev. {\bf B 64}, 045313 (2001).



\bibitem{Timofeev} A.~V. Larionov, V.~B. Timofeev, J. Hvam, and K.
Soerensen, ZhETF {\bf 117}, 1255 (2000) [JETP {\bf 90}, 1093
(2000)]; A.~V. Larionov and V.~B. Timofeev, Pis'ma ZhETF {\bf 73},
342 (2001) [JETP Lett. {\bf 73}, 301 (2001)].


\bibitem{Zrenner} T. Fukuzawa, E.~E. Mendez and J.~M. Hong,
    Phys. Rev. Lett. {\bf 64}, 3066 (1990);
J.~A. Kash, M. Zachau, E.~E. Mendez,   J.~M. Hong and T. Fukuzawa,
    Phys. Rev. Lett., {\bf 66},  2247 (1991).


\bibitem{Sivan}
U. Sivan, P.~M. Solomon and H. Shtrikman, Phys. Rev. Lett. {\bf
68}, 1196 (1992).


\bibitem{Snoke} For a general review of experiments in quantum wells,
see chapter 10 of S.~A. Moskalenko and D.~W. Snoke, {\it
Bose-Einstein Condensation of Excitons and Biexcitons and Coherent
Nonlinear Optics with Excitons} (Cambridge University Press, New
York, 2000).

\bibitem{Kim} E. Kim and M.~H.~W. Chan, Nature {\bf 427}, 225 (2004).

\bibitem{Anderson} M.~H. Anderson, J.~R. Ensher, Science {\bf 269}, 198 (1995).

\bibitem{Ensher} J.~R. Ensher, D.~S. Jin, M.~R. Matthews, C.~E.
Wieman, and E.~A. Cornell, Phys.~Rev.~Lett.
{\bf 77}, 4984 (1996).

\bibitem{Ketterle} W. Ketterle and N.~J. Druten, Phys.~Rev.  {\bf A
54}, 656 (1996).

\bibitem{Ketterle_Miesner} W. Ketterle and H.-~J. Miesner, Phys.~Rev.  {\bf A
56}, 3291 (1997).

\bibitem{Daflovo} F. Daflovo, S. Giorgini and L.P. Pitaevskii,
Rev.~Mod.~Phys. {\bf 71}, 463
  (1999).

\bibitem{Lozovik_tbp} Yu.~E. Lozovik, to be published.

\bibitem{Pomirchy} Yu.~E. Lozovik and L.~M. Pomirchy,
Sol.~St.~Commun. {\bf 89}, 145
(1994).

\bibitem{Ruvinsky_ps} Yu.~E. Lozovik and A.~M. Ruvinsky, Physica
Scripta {\bf 58}, 90 (1998).

\bibitem{Ruvinsky_jetp} Yu.~E. Lozovik and A.~M. Ruvinsky, ZhETF
{\bf 114}, 1451 (1998) [JETP {\bf 87}, 788 (1998)].

\bibitem{Ablyazov} N.~N. Ablyazov, M.~E. Raikh, and A.~L. Efros, Fiz.
Tverd. Tela {\bf 25}, 353 (1983) [Sov. Phys. Solid State {\bf 25},
199 (1983)].

\bibitem{Gevorkyan} Zh.~S. Gevorkyan and Yu.~E. Lozovik, Fiz.
Tverd. Tela {\bf 27}, 1800 (1985) [Sov. Phys. Solid State {\bf
27}, 1079 (1985)].

\bibitem{Berman_Ruvinsky} Yu.~E. Lozovik, O.~L. Berman and A.~M.
Ruvinsky, Pis°Øma ZhETF {\bf 69}, 573 (1999) [JETP Lett.
{\bf 69}, 616 (1999)].


\bibitem{Yudson} Yu.~E. Lozovik and V.~I. Yudson, Physica  {\bf A 93},
493 (1978).

\bibitem{Abrikosov} A. A. Abrikosov, L. P. Gorkov and I. E.
Dzyaloshinski, {\em Methods of Quantum Field Theory in Statistical
Physics} (Prentice-Hall, Englewood Cliffs. N.J., 1963).



\bibitem{Kosterlitz} J.~M. Kosterlitz and D.~J. Thouless, J.~Phys. {\bf C 6},
   1181 (1973); D.~R. Nelson and J.~M. Kosterlitz,
    Phys. Rev. Lett. {\bf 39}, 1201 (1977).

\bibitem{Nishanov} Yu.~E. Lozovik and V.~N. Nishanov,  Fiz. Tverd. Tela {\bf
18}, 3267 (1976) [Sov. Phys. Solid State {\bf 18}, 1905 (1976)].


\bibitem{John_Stephen} S. John and M.~J. Stephen, Phys.~Rev.  {\bf B
28}, 6358 (1983).



\bibitem{Halperin} B.~I. Halperin and T.~M. Rice, Solid State Phys. {\bf 21},
115 (1968).



\bibitem{Keldysh} L.V.Keldysh and A.N.Kozlov, ZhETF {\bf 54}, 978 (1968)
[JETP {\bf 27}, 521 (1968)].

\bibitem{Huang} K. Huang, in
{\it Bose-Einstein Condensation}, A. Griffin, D. W. Snoke, S.
Stringari, Eds. (Cambridge Univ. Press, Cambridge, 1995), p.p.
31-50.

\bibitem{Huang_Meng} K. Huang and H.~F. Meng,
Phys. Rev. Lett. {\bf 69}, 644 (1992); H.~F. Meng, Phys. Rev. {\bf
B 49},  1205 (1994).

\bibitem{Lopatin} A.~V. Lopatin and V.~M. Vinokur, Phys. Rev. Lett.
{\bf 88}, 235503 (2002).

\bibitem{Fisher} M.~P.~A. Fisher, P.~B. Weichman, G. Grinstein,
and D.~S. Fisher, Phys. Rev. {\bf B 40},  546 (1989).

\bibitem{Mahan}  G.~D. Mahan, {\it Many-Particle Physics}, Plenum Press,
   New York (1990).

\bibitem{Snoke-apl} V. Negoita, D.W. Snoke, and K. Eberl, Applied Physics
Letters {\bf 75}, 2059 (1999).

\end{thebibliography}
\end{document}